\documentclass[conference,a4paper]{IEEEtran}
\usepackage{xcolor}
\usepackage{balance}
\usepackage{subfigure}
\usepackage{amsmath,amssymb,amsfonts}
\usepackage{float}

\IEEEoverridecommandlockouts
\usepackage[left=1.4cm,right=1.4cm,top=1.8cm, bottom =4.4cm]{geometry}

\ifCLASSINFOpdf
  \usepackage[pdftex]{graphicx}
\else
  \usepackage[dvips]{graphicx}
\fi
\ifCLASSOPTIONcompsoc
\usepackage[caption=false,font=normalsize,labelfont=sf,textfont=sf]{subfig}
\else
 \usepackage[caption=false,font=footnotesize]{subfig}
\fi
\begin{document}
%
\title{Joint Beamforming for Multi-user Multi-target FD ISAC System: A Hybrid GRQ-GA Approach}

\author{\IEEEauthorblockN{
Duc Nguyen Dao\IEEEauthorrefmark{1},   
Haibin Zhang\IEEEauthorrefmark{2},   
André B. J. Kokkeler\IEEEauthorrefmark{1},    
Yang Miao\IEEEauthorrefmark{1}      
}                                     
\IEEEauthorblockA{\IEEEauthorrefmark{1}
Radio Systems Group, University of Twente, Enschede, The Netherlands, \{d.d.n.dao, a.b.j.kokkeler, y.miao\}@utwente.nl}
\IEEEauthorblockA{\IEEEauthorrefmark{2}
Department of Networks, TNO, The Hague, The Netherlands, haibin.zhang@tno.nl}
 \IEEEauthorblockA{}
}
\maketitle
\begin{abstract}
In this paper, we consider a full-duplex (FD) Integrated Sensing and Communication (ISAC) system, in which the base station (BS) performs downlink and uplink communications with multiple users while simultaneously sensing multiple targets. In the scope of this work, we assume a narrowband and static scenario, aiming to focus on the beamforming and power allocation strategies. We propose a joint beamforming strategy for designing transmit and receive beamformer vectors at the BS. The optimization problem aims to maximize the communication sum-rate, which is critical for ensuring high-quality service to users, while also maintaining accurate sensing performance for detection tasks and adhering to maximum power constraints for efficient resource usage. The optimal receive beamformers are first derived using a closed-form Generalized Rayleigh Quotient (GRQ) solution, reducing the variables to be optimized. Then, the remaining problem is solved using floating-point Genetic Algorithms (GA). The numerical results show that the proposed GA-based solution demonstrates up to a 98\% enhancement in sum-rate compared to a baseline half-duplex ISAC system and provides better performance than a benchmark algorithm from the literature. Additionally, it offers insights into sensing performance effects on beam patterns as well as communication-sensing trade-offs in multi-target scenarios.
\end{abstract}

\vskip0.5\baselineskip
\begin{IEEEkeywords}
Integrated Sensing and Communication (ISAC), full-duplex (FD), joint beamforming, Generalized Rayleigh Quotient (GRQ), Genetic Algorithm (GA).
\end{IEEEkeywords}

%

\section{Introduction}
Integrated Sensing and Communication (ISAC) is a key technique for 6G networks, combining sensing and communication in a unified system to enhance both functions \cite{b1}. Designing an ISAC system involves addressing the challenge of resource allocation across dimensions such as time, frequency, and/or spatial domain \cite{b1}. In this paper, we focus on spatial resource allocation. 

The multiple-input multiple-output (MIMO) technique plays a key role in ISAC, utilizing spatial degrees of freedom (DoF) to perform multi-target sensing and multi-user communication by synthesizing beams towards users and sensing targets. In \cite{b3}, a joint beamforming for separate communication and sensing waveforms was proposed, which provides extra DoF and enhances sensing accuracy. However, all of these works so far only considered the coexistence of radar sensing and downlink communication signals. In practice, it might be desired that a BS may serve multiple users simultaneously in both downlink and uplink, i.e. full-duplex (FD) transmission, for higher resource efficiency. While \cite{b4} explored FD ISAC joint beamforming for downlink and uplink sum-rate optimization, only one sensing target was considered. In practical applications, scenarios involving simultaneous detection of multiple targets near multiple users are highly likely, and addressing these challenges will enhance the system's integration and operational efficiency. The primary challenge of this is mitigating interference between target echos  and the likely correlation between sensing and communication channels. 

 Furthermore, most resource allocation problems in ISAC are non-convex due to the associated objective functions and constraints, making them challenging to solve. In the literature, these non-convex problems are often approximated as convex problems using various relaxation techniques, such as  semi-definite relaxation (SDR) and function approximation such as successive convex approximation (SCA) \cite{b4, b5}. However, while these methods simplify the original problems, they may lose some accuracy due to the approximations and matrices decomposition. On the other hand, Genetic Algorithms (GAs) \cite{b6} have been explored as an alternative approach in some studies as they can handle non-convex problems directly without requiring convex approximations. What sets our approach apart is the use of a GA combined with a Generalized Rayleigh Quotient (GRQ) optimization to reduce the complexity of solving resource allocation problems, which will be detailed in Section \ref{sec3}. This combination represents a novel contribution compared to existing methods that typically rely solely on relaxation techniques.
 
Motivated by the challenges of FD ISAC and the complexity of the corresponding optimization problems, this paper studies an FD ISAC system where the base station (BS) transmits signals to multiple downlink users and receives signals from multiple uplink users, while simultaneously detecting multiple targets. The objective is to maximize the total sum-rate of FD communication (both downlink and uplink) under the sensing signal-to-interference-plus-noise ratio (SINR) constraint for reliable sensing performance. This balance is crucial in applications requiring both high communication efficiency and accurate sensing, such as autonomous systems and smart environments \cite{b1}. The paper contributes a joint beamforming strategy for the FD ISAC system, including managing interference between uplink and downlink signals, integrating communication and sensing tasks, and effectively allocating power among all functions. A hybrid GRQ-GA approach is employed to derive solutions.Numerical results demonstrate the method's performance benefits over benchmark schemes, provide insights into communication-sensing trade-offs, and highlight the impacts of sensing requirements on beam pattern. In the scope of this work, we focus on a narrowband, static scenario and limit our sensing task to target detection, assumi ng a monostatic radar integrated with BS. These assumptions make the joint beamforming problem more tractable while still being relevant to many practical applications, such as radar-based object detection in controlled environments \cite{b1}. 

$Notations$: Matrices are represented by bold uppercase letters, vectors by bold lowercase letters, and scalars by regular font. We use $(.)^T$ and $(.)^H$ to denote the transpose and the Hermitian transpose, respectively. $\boldsymbol{I}_N$ stands for identity matrix of size $N\times N$, and $[.]_{a,b}$ denotes the $(a,b)$-th entry of a matrix. The imaginary unit is represented as $i^2 = -1$, and $\odot$  denotes the element-wise poroduct of two vectors.

\section{System model and Problem formulation}\label{sec2}
\subsection{System model}
\begin{figure}[t]
\centerline{\includegraphics[width=4.5 cm]{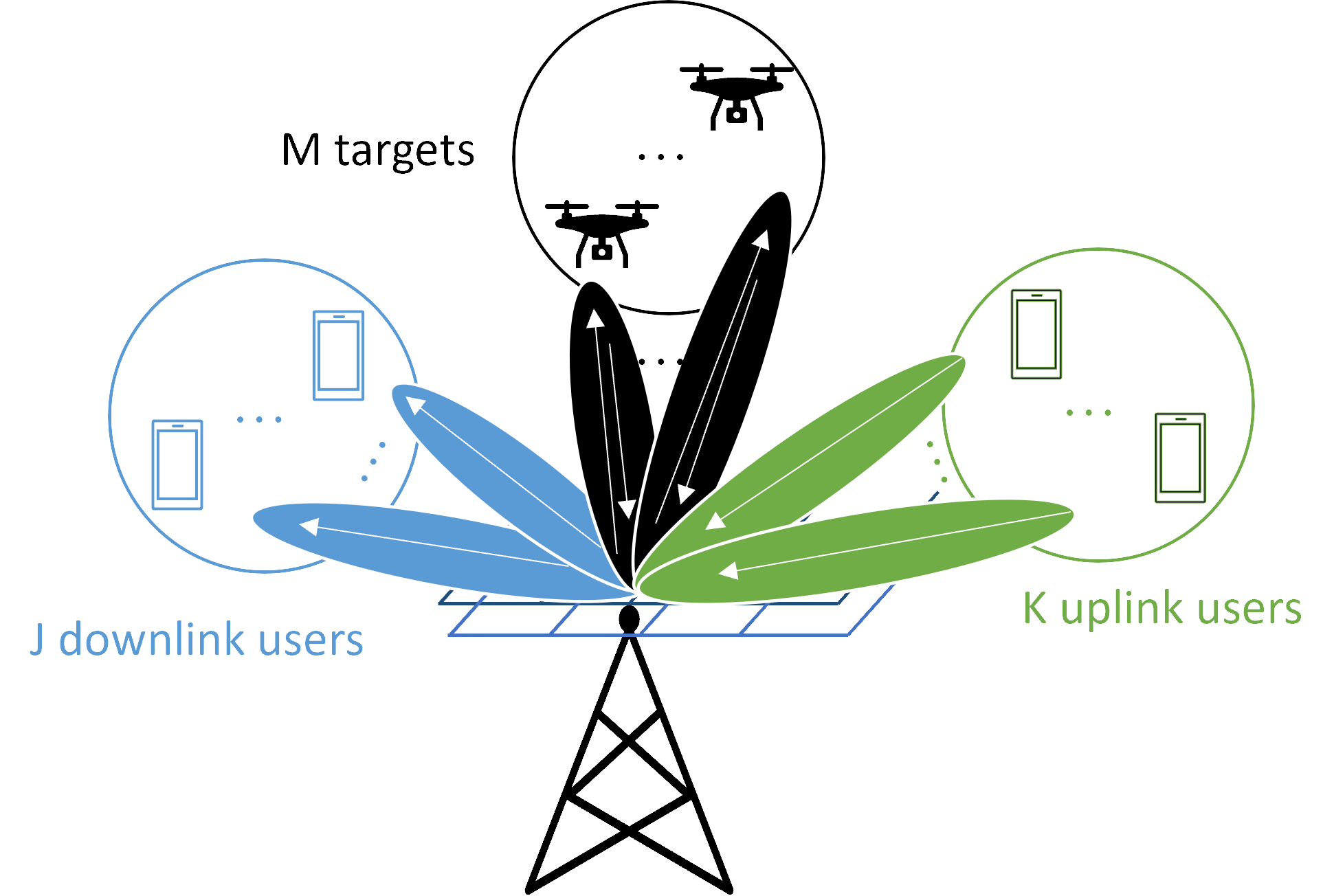}}
\caption{ISAC-based scenario: for illustration purposes, different types of users and targets are grouped and spatially separated. In practical scenarios, their locations may vary randomly.}
\vspace{-10pt}
\label{system_model}
\vspace{-12pt}
\end{figure}

We consider a MIMO ISAC system as shown in Fig. \ref{system_model}, where the BS has two separate antenna arrays: a transmit array with $N_t$ elements and a receive array with $N_r$ elements. In this setup, $J$ represents the number of downlink users, $K$ denotes the number of uplink users, and $M$ indicates the number of sensing targets. The BS performs as both transmitter and receiver. As a transmitter, the BS array sends downlink signals to $J$ single-antenna users while simultaneously performing target detection on $M$ targets using dedicated sensing signals. As a receiver, the BS collects communication signals from $K$ uplink users as well as echo signals from targets for sensing. The transmit and receive arrays are positioned closely to one another, which means that the angles of the targets observed by the receiver and transmitter are identical.

In this work, we assume prior knowledge that all users' locations are known via the downlink synchronization process compliant with the 5G NR protocol. It is also assumed that the target positions are roughly estimated during beam scanning in the downlink synchronization phase with additional dedicated radar receiver integrated in the BS. 
After the aforementioned synchronization phase, the proposed optimization produces multiple transmit beams for simultaneous downlink communications and illuminating the target, as well as multiple receive beams for simultaneous uplink communications and receiving target echos.

First, let the joint transmit signal $\boldsymbol{x} \in \mathbb{C}^{N_t \times 1}$ be sent from the BS for simultaneous sensing and downlink multi-user communication given by:
\begin{equation}
    \boldsymbol{x}=\sum_{j=1}^{J} \boldsymbol{v}_{c, j} c_j+\sum_{m=1}^M \boldsymbol{v}_{s, m} s_m,
\end{equation}
where $\boldsymbol{v}_{c,j} \in \mathbb{C}^{N_t \times 1}$ and $\boldsymbol{v}_{s,m}\in \mathbb{C}^{N_t \times 1}$ are the transmit beamforming vectors towards downlink user $j$ and target $m$, respectively. Let $c_j$ be the dedicated downlink data symbol for user $j$ and $s_m$ denote the dedicated sensing signal for target $m$. The data symbols $c$ and $s$ are assumed to have unit power $\mathbb{E}{\{|c_j|^2}\} = 1, \forall j, \mathbb{E}{\{|s_m|^2}\} = 1, \forall m,$  and are independent with each other. The total BS transmit power is as follow:
\begin{equation}
    P_{\text{tx}} = \sum_{j=1}^J\left\|\boldsymbol{v}_{c,j}\right\|^2 + \sum_{m=1}^M\left\|\boldsymbol{v}_{s,m}\right\|^2.
\end{equation}
Let $\boldsymbol{a}(\theta, \phi) \in \mathbb{C}^{N \times 1}$ be the steering vector of the array with $N$ elements corresponding to the angle ($\theta, \phi$)
\begin{equation}
    \boldsymbol{a}(\theta,\phi) = \frac{1}{\sqrt{N}}[\exp(i\boldsymbol{k}(\theta,\phi)\boldsymbol{u}_1),...,\exp(i\boldsymbol{k}(\theta,\phi)\boldsymbol{u}_n)]^T,
\end{equation}
where $\boldsymbol{k}(\theta,\phi) = \frac{2\pi}{\lambda_c}[\sin(\theta)\cos(\phi), \sin(\theta)\sin(\phi),\cos(\theta)]^T$ describes the phase variation, $\lambda_c$ is the carrier wavelength,  and $\boldsymbol{u}_n$ is the location vector of the $n$-th antenna element. 

In real-world scenario, communication and sensing channels are likely to be correlated. For example, radar targets may act as clusters in communication channels. In these cases, it is necessary to have a joint model to represent both communication and sensing channels. Denote $\boldsymbol{h}^{(c)}_{DL,j} \in \mathbb{C}^{N_t \times 1}$ as the channel between the downlink user $j$-th and the BS. For simplicity, we consider an ideal case where the sensing targets are treated as clusters within the communication channel, without the presence of additional background clusters. The channel can be mathematically expressed as:
\begin{equation}
\begin{aligned} 
  &  \boldsymbol{h}_{j} = \sqrt{N_t}(\frac{\lambda_c}{4\pi d_{bs,j}}\exp(\frac{-i2\pi d_{bs,j}}{\lambda_c})\boldsymbol{a}^{(tx)}_j +\\ &  \frac{\lambda_c}{(4\pi)^{3/2} d_{bs,m} d_{m,j}} \sum_{m=1}^M \sigma_m \exp(\frac{-i2\pi (d_{bs,m} + d_{bs,j})}{\lambda_c})\boldsymbol{a}^{(tx)}_m),
  \end{aligned}
\end{equation}
where $d_{bs,j}$, $d_{bs,m}$ are the distance between the BS and user $j$-th and target $m$-th, $\boldsymbol{a}^{(tx)}_j$ and $\boldsymbol{a}^{(tx)}_m \in \mathbb{C}^{N_t \times 1}$ are the transmit steering vectors to the direction of the $j$-th user and $m$-th target, respectively. $\sigma_m$ represents radar cross section (RCS) of the $m$-th target. The received signal at user $j$ is represented as
\begin{equation}
    y_{\text{DL},j}= \boldsymbol{h}_j^{H} \boldsymbol{v}_{c, j} c_j + \sum_{j^{\prime}=1, j^{\prime} \neq j}^J \boldsymbol{h}_j^{H} \boldsymbol{v}_{c, j'} c_{j'} + \sum_{m=1}^M \boldsymbol{h}_j^{H} \boldsymbol{v}_{s, m} s_m + n_j.
\end{equation}
This received downlink signal consists of four parts. The first part corresponds to the desired signal aimed to user $j$. The second and third part represent the interference from other downlink users and the sensing signals, respectively. The last component is the additive white Gaussian noise (AWGN) at user $j$ with the variance of $\sigma_j^2$. 

Similarly, the uplink channel $\boldsymbol{g}^{(c),LoS}_{UL,k} \in \mathbb{C}^{N_r \times 1}$ between $k$-th user and the BS is modeled as: 
\begin{equation}
\begin{aligned}
    &\boldsymbol{g}_k = \sqrt{N_r}(\frac{\lambda_c}{4\pi d_{bs,k}}\exp(\frac{-i2\pi d_{bs,k}}{\lambda_c})\boldsymbol{a}^{(rx)}_k + \\& \frac{\lambda_c}{(4\pi)^{3/2} d_{m,k} d_{bs,m}} \sum_{m=1}^M \sigma_m \exp(\frac{-i2\pi (d_{m,k} + d_{bs,m})}{\lambda_c})\boldsymbol{a}^{(rx)}_m),  
\end{aligned}
\end{equation}
where $d_{bs,k}$ and $\boldsymbol{a}^{(rx)}_k$ are the distance from the BS and the receiving steering vector of the $k$-th uplink user, respectively.

Let $t_k$ represent the uplink symbol transmitted from user $k$. We assume $\mathbb{E}{\{|t_k|^2}\} = 1$. The average power of transmit signal from user $k$ is denoted as $e_k$. Assume that the initial target's location is already estimated and known to the ISAC system. By combining the uplink signals, sensing reflected signals, and the SI signal, all of which are assumed to be uncorrelated with each other, the received signal at the BS can be expressed as follows:
\begin{equation}
    \boldsymbol{y} = \sum_{k=1}^K \boldsymbol{g}_k \sqrt{e_k} t_k + \sum_{m=1}^M \alpha_m \boldsymbol{A}_m \boldsymbol{x} + \boldsymbol{H}_{\text{SI}} \boldsymbol{x}  + \boldsymbol{n}_{BS}, 
\end{equation}

where $\boldsymbol{A}_m = \boldsymbol{a}^{(rx)}_m\boldsymbol{a}^{(tx)H}_m$ with $\boldsymbol{a}^{(rx)}_m$ and $\boldsymbol{a}^{(tx)}_m$ are the receive and transmit beamforming vectors toward the $m$-th user, respectively. $\alpha_m$ is the complex gain of the $m$-th target, consisting of antenna gains, two-way path loss, and RCS, which can be expressed as follows: 

\begin{equation}
     \alpha_m = \sqrt{N_t N_r}\frac{\lambda_c \sigma_m}{(4 \pi)^{3/2}{d_{bs,m}}^2}\exp(\frac{-i2\pi (2d_{bs,m})}{\lambda_c}),
\end{equation}

The sensing reflected signals from different targets and received signals from uplink users are assumed to be uncorrelated with each other.
$\boldsymbol{H}_{\text{SI}} \in \mathbb{C}^{N_r \times N_t}$ denotes the self-interference channel (SI) from the transmitter to the receiver of the BS. The SI channel is modeled following the approach of \cite{b4} as
\begin{equation}
    [H_{\text{SI}}]_{a,b} = \sqrt{\eta_{a,b}}\exp(\frac{-i2\pi d_{a,b}}{\lambda_c} ),
\end{equation}
 where $\eta_{a,b}$ and $d_{a,b}$ are the channel gain and distance between the $a$-th receive antenna element and the $b$-th transmit antenna element, respectively.

\subsubsection{Performance metrics}
The downlink communication, the SINR of the user $j$ is
\begin{equation}\label{eq4}
\gamma_{\text{DL},j}=\frac{\left|\boldsymbol{h}_j^H \boldsymbol{v}_{c,j}\right|^2}{\sum_{j^{\prime}=1, j^{\prime} \neq j}^J|\boldsymbol{h}_j^H \boldsymbol{v}_{c,j^{\prime}}|^2+\sum_{m=1} ^M|\boldsymbol{h}_j^H \boldsymbol{v}_{s,m}|^2+\sigma_j^2},
\end{equation}

For the received signal in the BS, a receive beamformer $\boldsymbol{w}_{s,m} \in \mathbb{C}^{N_r \times 1}$ is applied to restore the desired target sensing signal $m$. Therefore, the SINR of sensing target $m$ is calculated as
\begin{equation}\label{eq5}
\begin{aligned}
&\gamma_{\text{rad}, m}= \\
& \frac{|\boldsymbol{w}_{s, m}^H \alpha_m\boldsymbol{A}_m \boldsymbol{v}_{s,m}|^2}{\boldsymbol{w}_{s,m}^H(\sum _{\widehat{RAD}} + \sum_{\text{UL}} + \sum_{\text{DL}}+ \xi +\sigma_{BS}^2 \boldsymbol{I}_{N_r}) \boldsymbol{w}_{s, m}},  
\end{aligned}
\end{equation}
where $\sum _{\widehat{RAD}} = \sum _{m^{\prime}=1, m^{\prime} \neq m}^M |\alpha_{m^{\prime}}\boldsymbol{A}(\theta_{m^{\prime}}, \phi_{m^{\prime}}) \boldsymbol{v}_{s,{m^{\prime}}}|^2$ represents the summation of sensing echoes from other $m^{\prime}$ targets. $\sum_{\text{UL}} = \sum_{k=1}^K e_k \boldsymbol{g}_k \boldsymbol{g}_k^H$ denotes the interference from uplink transmission of all $K$ users, $\sum_{\text{DL}} = \sum _{m=1}^M |\alpha_{m}\boldsymbol{A}(\theta_{m}, \phi_{m})\sum _{j=1}^J \boldsymbol{v}_{c,j}|^2$ represents the interference of downlink signals reflected by all $M$ targets, and $\xi = |\boldsymbol{H}_{\text{SI}}\boldsymbol{x}|^2$ indicates the self-interference in FD BS.

Similarly, to obtain the dedicated uplink signal of the user $k$, we apply the receive beamformers $\boldsymbol{w}_{c,k} \in \mathbb{C}^{N_r \times 1}$ to the received signal in the BS. We derive the SINR of uplink $k$-th user as
\begin{equation}
\begin{aligned}\label{eq6}
& \gamma_{\text{UL}, k}=  \frac{\left|\boldsymbol{w}_{c, k}^H \boldsymbol{g}_{k}t_{k}\right|^2}{\boldsymbol{w}_{c,k}^H(\sum _{\widehat{UL}}+\sum _{\text{rad}} + \sum_{\text{DL}} + \xi + \sigma_{BS}^2 \boldsymbol{I}_{N_r}) \boldsymbol{w}_{c, k}},
\end{aligned}
\end{equation}
where the interference of other $k^\prime$ uplink transmissions is $\sum _{\widehat{UL}} = \sum _{k^{\prime}=1, k^{\prime} \neq k}^K e_k \boldsymbol{g}_{k^\prime} \boldsymbol{g}_{k^\prime}^H $, and $\sum _{\text{rad}} = \sum _{m=1}^M |\alpha_{m}\boldsymbol{A}(\theta_{m}, \phi_{m}) \boldsymbol{v}_{s,m}|^2$ represents the sumation of sensing interference reflected from all $M$ targets.

\subsection{Problem formulation}
We employ spectral efficiency for the evaluation of uplink and downlink communication performance. For the evaluation of sensing performance, we apply per-target SINR's, which could be further translated into detection probability assuming certain sensing waveforms \cite{b7}. Let $\tau_{\text{DL}}$ and $\tau_{\text{UL}}$ be the spectral efficiency of downlink and uplink communications, respectively, which can be represented as
\begin{equation}
    \tau_{\text{DL}} = \sum_{j=1}^J log_2 (1 + \gamma_{\text{DL},j}), \quad\tau_{\text{UL}} = \sum_{k=1}^K log_2 (1 + \gamma_{\text{UL},k}).
\end{equation}
 The objective is to maximize the spectral efficiency of both downlink and uplink communication, for which the problem will be formulated as follows:
\begin{subequations}\label{eq9}
    \begin{align}
    &\underset{\substack{\{\boldsymbol{v}_{c,j}\}, \{\boldsymbol{v}_{s,m}\}, \{e_k\} \\ \{\boldsymbol{w}_{c,k}\}, \{\boldsymbol{w}_{s,m}\}}}{\operatorname{maximize}}  \;\; \rho \tau_{\text{DL}} + (1-\rho) \tau_{\text{UL}}, \\
    \text { s.t \;\;} & \gamma_{\mathrm{rad,m}} \geq \mu_{\mathrm{rad,m}}, \quad \forall m,\\
    & \gamma_{\mathrm{DL,j}} \geq \mu_{\mathrm{DL,j}}, \forall j, \quad \gamma_{\mathrm{UL,k}} \geq \mu_{\mathrm{UL,k}}, \forall k,\\
    & P_{\text{tx}} \leq P_{\text{max}}, \quad e_k \leq P_k, \quad \forall k, \\
    & \left\|\boldsymbol{v}_{c,j}\right\|^2 \leq P_{0 }, \forall j, \quad\left\|\boldsymbol{v}_{s,m}\right\|^2 \leq P_{0 }, \forall m,
    \end{align}
\end{subequations}
where $\rho$ is the tradeoff parameter between downlink and uplink communication. Constraints (\ref{eq9}b) and (\ref{eq9}c) ensure the minimum SINR requirements for sensing and communication performance, where $\mu_{\text{rad}}$, $\mu_{\text{DL}}$ and $\mu_{\text{UL}}$ denote the minimum acceptable SINR for each target, downlink user and uplink user, respectively. Constraint (\ref{eq9}d) ensures the maximum transmit power and the power constraint of each uplink user, and 
(\ref{eq9}e) limits the maximum power per antenna element.

\section{Joint FD beamforming}\label{sec3}
In this section, we tackle the non-convex problem as shown in (\ref{eq9}). We first use the GRQ to find the closed-form solutions for obtaining optimal receive beamformers of sensing $\boldsymbol{w}_{s}$ and uplink communication $\boldsymbol{w}_{c}$. These results are then substituted back into the original problem (\ref{eq9}), which effectively frees the receive beamformer vectors from the optimized variables. Then, we solve the equivalent problem using GA.
\subsection{Closed-form solutions for receive beamformers}
Note that the receive beamformer vectors of sensing $\{\boldsymbol{w}_{s,m}\}$ does not affect the objective function of (\ref{eq9}) and $\boldsymbol{w}_{c,k}$ only has effects on $\gamma_{\text{UL},k}$. First, let us rewrite the sensing SINR of the $m$-th target (\ref{eq5}) as
\begin{equation}\label{eq12}
\begin{aligned} 
\gamma_{\text{rad}, m} = &\left|\alpha_m\right|^2 \boldsymbol{a}^{(rx)H}_m \boldsymbol{v}_{s}\boldsymbol{v}^H_{s} \boldsymbol{a}^{(tx)}_m \\
& \times \frac{\boldsymbol{w}_{s,m}^{{H}} \boldsymbol{a}^{(rx)}_m \boldsymbol{a}^{(rx) H}_m \boldsymbol{w}_{s,m}}{\boldsymbol{w}_{s,m}^{H} \boldsymbol{D} \boldsymbol{w}_{s,m}} , \\ 
\end{aligned}
\end{equation}
 where $\boldsymbol{D} = (\sum _{\widehat{RAD}} + \sum_{\text{UL}} + \sum_{\text{DL}}+ \xi +\sigma_{BS}^2 \boldsymbol{I}_{N_r})$ and $\boldsymbol{E} = (\sum _{\widehat{UL}}+\sum _{\text{rad}} + \sum_{\text{DL}} + \xi + \sigma_{BS}^2 \boldsymbol{I}_{N_r})$. The uplink SINR of the $k$-th user in (\ref{eq6}) is
\begin{equation}\label{eq13}
\begin{aligned}
\gamma_{\text{UL}, k} = &e_k \times \frac{\boldsymbol{w}_{c,k}^{H} \boldsymbol{g}_k \boldsymbol{g}_k^{H}\boldsymbol{w}_{c,k}}{\boldsymbol{w}_{c,k}^{H} \boldsymbol{E} \boldsymbol{w}_{c,k}}, \\ 
\end{aligned}
\end{equation}

 It can be observed that the problems in (\ref{eq12}) and (\ref{eq13}) are instances of the GRQ. As shown in \cite{b10}, $\gamma_{\text{rad}, m}$ and $\gamma_{\text{UL}, k}$ can achieve their maximum values with the optimal solutions for $\{\boldsymbol{w}_{s,m}\}$ and $\{\boldsymbol{w}_{c,k}\}$, respectively, as follows: 

\begin{equation}\label{eq14}
    \boldsymbol{w}_{s,m}^* = \boldsymbol{D}^{-1} \boldsymbol{a}^{(rx)}_m, \quad \boldsymbol{w}_{c,k}^* = \boldsymbol{E}^{-1} \boldsymbol{g}_k,
\end{equation}
Substituting the optimal solutions of $\boldsymbol{w}_{s,m}^*$ and $\boldsymbol{w}_{c,k}^*$ obtained in (\ref{eq14}) into (\ref{eq12}) and (\ref{eq13}), we come to the following optimal SINR expressions: 
\begin{equation}
\begin{aligned}
\gamma_{\text{rad}, m} = &\left|\alpha_m\right|^2 \boldsymbol{a}^{(tx)H}_m \boldsymbol{v}_{s,m}\boldsymbol{v}^H_{s,m} \boldsymbol{a}^{(tx)}_m \\
& \times \boldsymbol{a}^{(rx)H}_m\boldsymbol{D}^{-1} \boldsymbol{a}^{(rx)}_m,  
\end{aligned}
\end{equation}
\begin{equation}
\begin{aligned}
\gamma_{\text{UL}, k} = &e_k \boldsymbol{g}^H_k\boldsymbol{E}^{-1} \boldsymbol{g}_k, 
\end{aligned}
\end{equation}
\subsection{Floating point Genetic Algorithm-based optimization}
GA is an optimization algorithm inspired by the principles of natural selection and biological evolution \cite{b6}. It evolves a population of potential solutions through three main processes: selection, crossover, and mutation. The floating-point GA is more efficient for continuous variables than the standard GA, which relies on binary encoding and decoding. Additionally, the floating-point GA offers higher precision by searching the entire solution space rather than just discrete points, as with the standard GA.

Since a GA can only deal with real variables, the transmit beamformer vectors $\boldsymbol{v}_c$ and $\boldsymbol{v}_s$ will first be decomposed into amplitude parts and phase parts:
\begin{equation}
\begin{split} &
    \boldsymbol{v}_{c,j} = \boldsymbol{\psi}_{c,j}\odot \exp(i\boldsymbol{\beta}_{c,j}) , \forall j\in J,\\&
    \boldsymbol{v}_{s,m} = \boldsymbol{\psi}_{s,m}\odot \exp(i\boldsymbol{\beta}_{s,m}) , \forall m\in M,
    \end{split}
\end{equation}
where $\boldsymbol{\psi}_{c,j}, \boldsymbol{\psi}_{s,m} \in \mathbb{C}^{N_t \times 1}$, and $\boldsymbol{\beta}_{c.j}, \boldsymbol{\beta}_{s,m} \in \mathbb{C}^{N_t \times 1}$ represent the amplitudes and phases of transmit beamforming vectors of downlink user $j$ and sensing target $m$, respectively. To initiate the population, each gene will be expressed by a floating-point number within an interval. For the phase of each transmit beamformer, the interval will be taken as [0, 2$\pi$), and for the amplitudes, the interval will be [0, $\sqrt{P_0}$]. Lastly, the interval of uplink power $\boldsymbol{e}_k$ will be [0, $P_{k}$], where $P_0, P_k$ is maximum power of per-antenna and uplink transmission. The population of candidate solutions is randomly initialized within the predefined range. 

To evaluate the performance of each candidate solution against the optimization objective, the fitness score is calculated as in (\ref{eq9}) at each iteration, with constraints addressed using penalty functions \cite{b11}. According to \cite{b12}, binary tournament selection offers lower time complexity while achieving similar performance to linear ranking selection. Two random solutions are selected from the population, with the candidate having the higher fitness score chosen for reproduction and the lower fitness score discarded. 

In a floating-point GA, new variables in the offspring are generated by combining two parent solutions. Denote $p_{m,n}$ and $p_{f,n}$ as the $n$-th variable in the mother and father chromosome, respectively. The value of the new single offspring variable, $p_{new}$, is created by performing scattered crossover:
\begin{equation}
    p_{new} = \zeta p_{m,n} + (1-\zeta) p_{f,n},
\end{equation}
where $\zeta$ is a random number on the interval [0, 1]. To perform the mutation operator, a random element $I_n$ of a random parent is selected and replaced with a new random variable within the specified lower and upper bounds.

The overall complexity of algorithm is approximately $\mathcal{O}\Big(I_{max} N_{max}((M+J)N_t+K)((K+M)N_r^3+KN_r^2+(J+M)N_t^2+(M+K)N_tN_r)\Big)\approx \mathcal{O}\Big(I_{max} N_{max} (M+J)(M+K)N_tN_r^3\Big)$
 if $N_t, N_r \gg J, K, M$, and $I_{max}$ is the number of iteration, $N_{max}$ is the population size. With the number of receive antennas $N_r$ increases, the algorithm's complexity grows rapidly.
\subsection{Baseline scenarios}
To show the advantages of the proposed GA-based approach, we consider the following two baseline scenarios:
\subsubsection{Communication-Only scenario} 
In this baseline scenario, an FD communication-only system is considered. As there are no sensing functionalities, this baseline scenario can help to evaluate how effective the communication and sensing integration is and serve as the upperbound for the communication performance. The corresponding algorithms for this FD communication system were proposed in \cite{b13}.
\subsubsection{ISAC with TDD HD communication}
To show the advantages of FD system, we consider a time-division duplex (TDD) system for comparison, where downlink and uplink communications are scheduled in separated time slots, while the BS continuously performs sensing function with separate sensing signals. This implies that the (downlink) sensing signals coexists with either downlink communication signal or uplink communication signal, but not both simultaneously. The communication performance will be average of those achieved in two different types of time slots (uplink communication plus sensing, and downlink communication plus sensing). The optimization problems can be solved using methods in \cite{b13} with some necessary modifications. 
\subsubsection{SDR-SCA} 
The algorithm adopted in \cite{b4} is considered as a benchmark for FD ISAC with simultaneous up/down-links and sensing. It is important to note that the SCA-SDR algorithm did not investigate the multi-target scenarios.
\section{Numerical results}\label{sec4}
\begin{table}[t]
\caption{Genetic Algorithm setup parameters}
\begin{center}
\begin{tabular}{cc}
\hline
\textbf{Parameter} & \textbf{Value} \\
\hline
Population size & 200 \\
Elite count & 10  \\
Crossover ratio & 0.8  \\
Maximum generation & 50  \\
\hline
\vspace{-25pt}
\end{tabular}
\label{tab1}
\end{center}
\end{table}
As reported in \cite{b14}, the circular array antenna achieved better spatial resolution, a narrower beam, and deeper nulls compared to the planar array with the same number of elements. Therefore, for a better beamforming performance, we choose the circular array antenna as the antenna model in simulations with Nt = Nr = 32 elements for its transmitter and receiver. The BS serves $J = 2$ downlink users, $K = 2$ uplink users and the number of sensed target M is varied from 1 to 4. The system operates at 39 GHz. To have a flat fading channel \cite{b15}, the bandwidth is set to 10 MHz which corresponds to a noise power of $-103.78$~dBm at room temperature. For simplicity, we assume a noise figure of 9 dB \cite{b16} to all users, and the RCS of 0 $dBm^2$ for all targets. Additionally, all DL users are positioned 250 m from the BS, UL users are located 200 m away, and targets are situated 100 m from the BS. The tradeoff parameter is set to $\rho$ = 0.5, representing that downlink and uplink communication have the same weight. The power budget of the BS, per antenna power, and maximum power of each uplink user is considered as $P_{max} = 17$~dBW, $P_0 = P_k = 1$~dBW, respectively \cite{b9}. The simulation parameters for the GA are presented in Table \ref{tab1}. The minimum SINR's of downlink and uplink communications are set to $\mu_{\text{DL},j} = 12$~dB $\forall j$, $\mu_{\text{UL},k} = 10$~dB, $\forall k$ \cite{b4}, and the minimum SINR of the sensing tasks $\mu_{\text{rad},m}$ is varying from $14$ to $20$~dB, $\forall m$. Results are averaged over 200 Monte Carlo simulations to address the randomness of GA as a stochastic optimizer.
\begin{figure}[t]
\centerline{\includegraphics[width=6 cm]{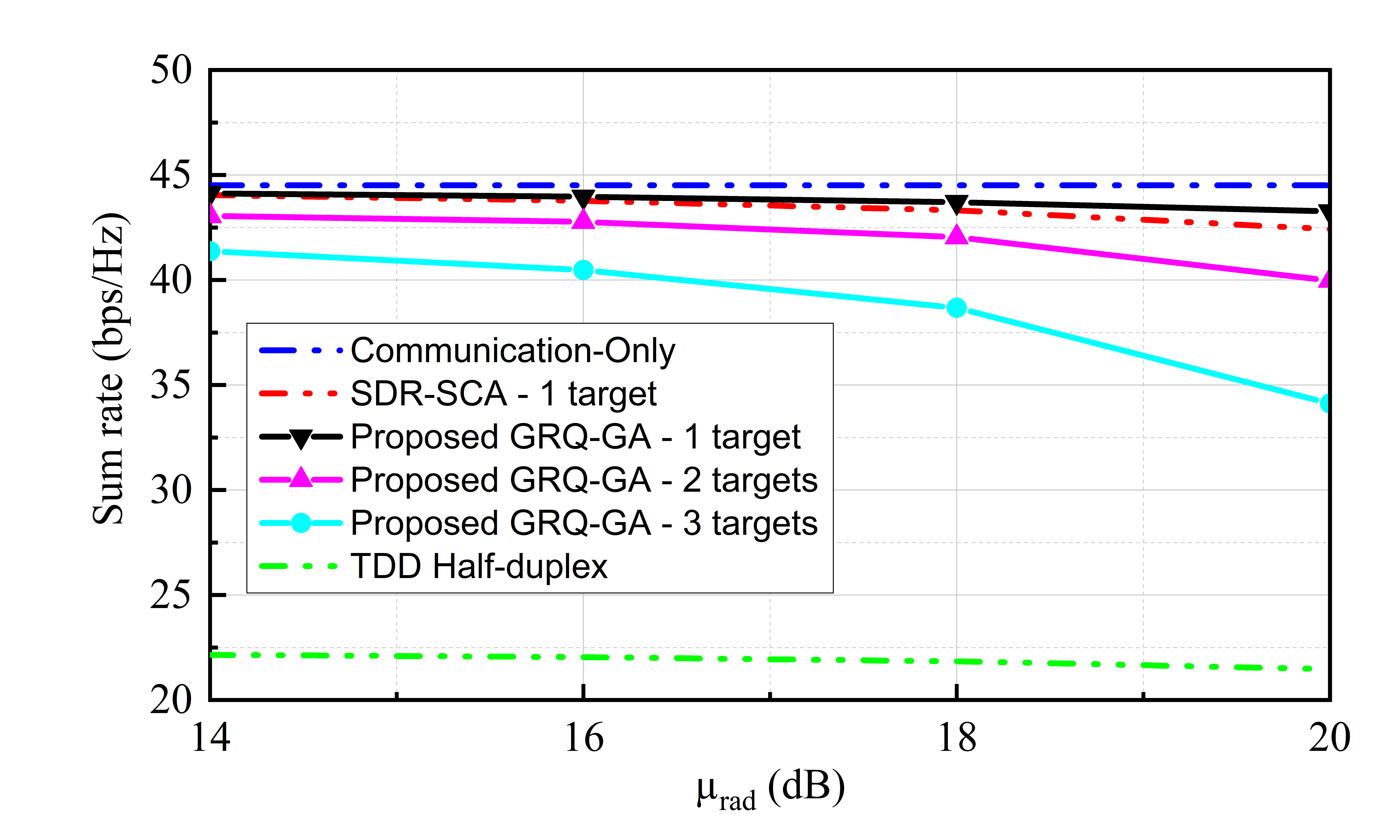}}
\caption{Multi-user capacity versus SINR threshold $\mu_{\text{rad}}$ with J = 2, K = 2. }
\label{capacity}
\end{figure}
 \vspace{-30pt}
\begin{figure}[H] 
    \subfigure[Azimuth plane $\theta = 20^\circ$]{\includegraphics[width=4.2 cm]{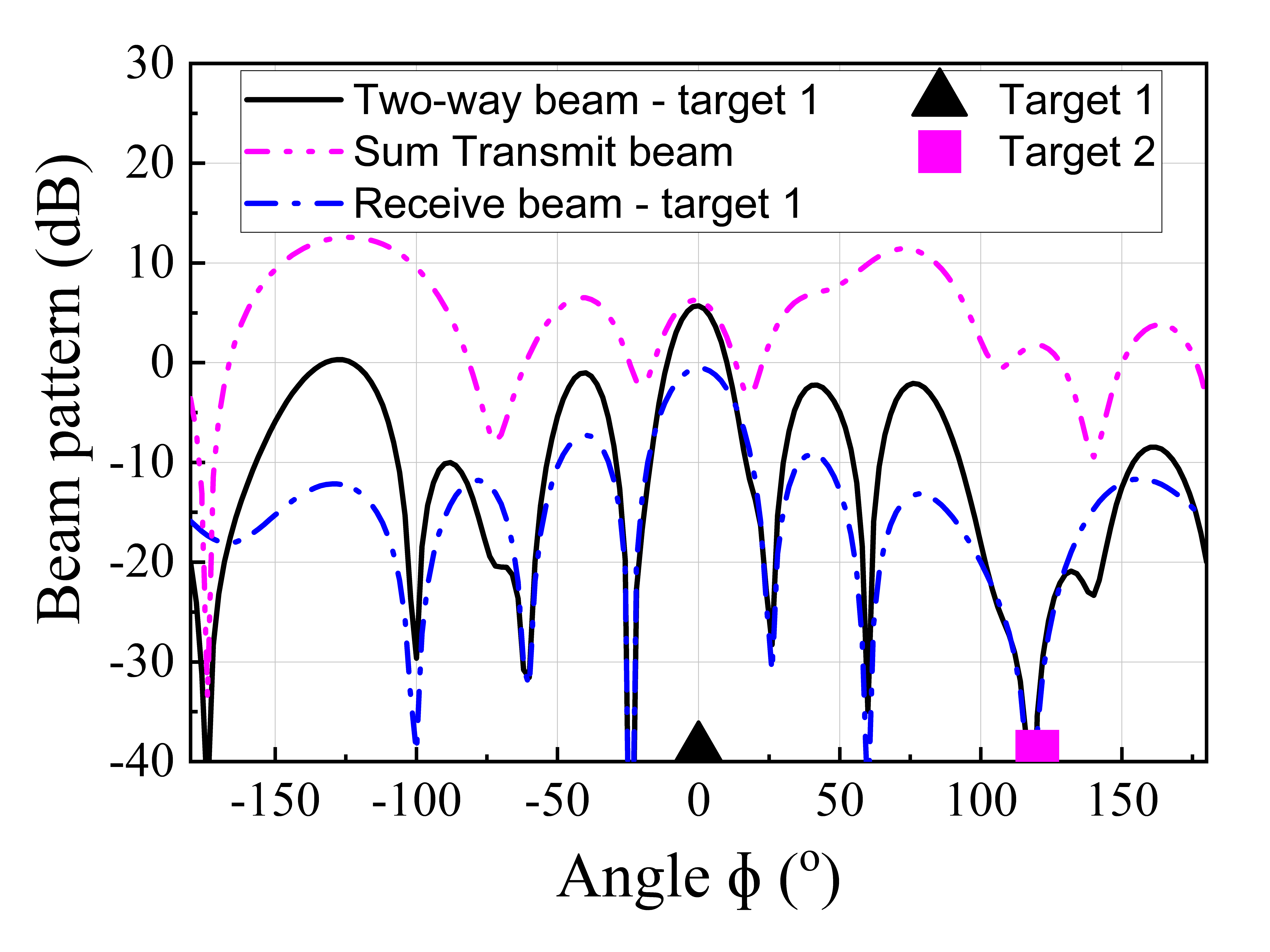}}
    \hfil
    \subfigure[Elevation plane $\phi = 0^\circ$]{\includegraphics[width=4.2 cm]{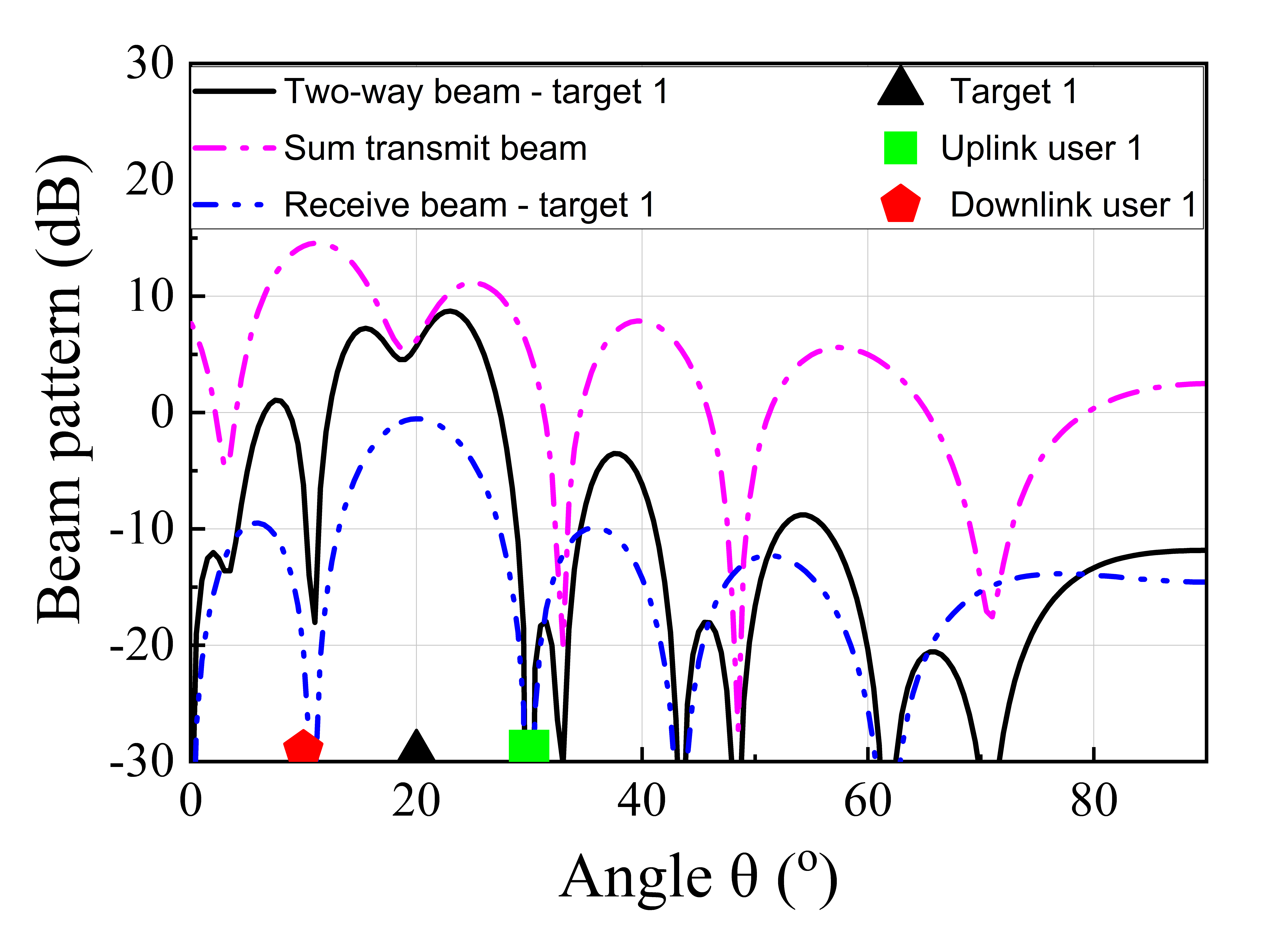}}
    \caption{Beam pattern of target 1 with required SINR $\mu_{\text{rad}}$ = 14 dB.}
    \vspace{-10pt}
    \label{beam_pattern1}
\end{figure}
\vspace{-10pt}
\begin{figure}[H] 
    \subfigure[Azimuth plane $\theta = 20^\circ$]{\includegraphics[width=4.2 cm]{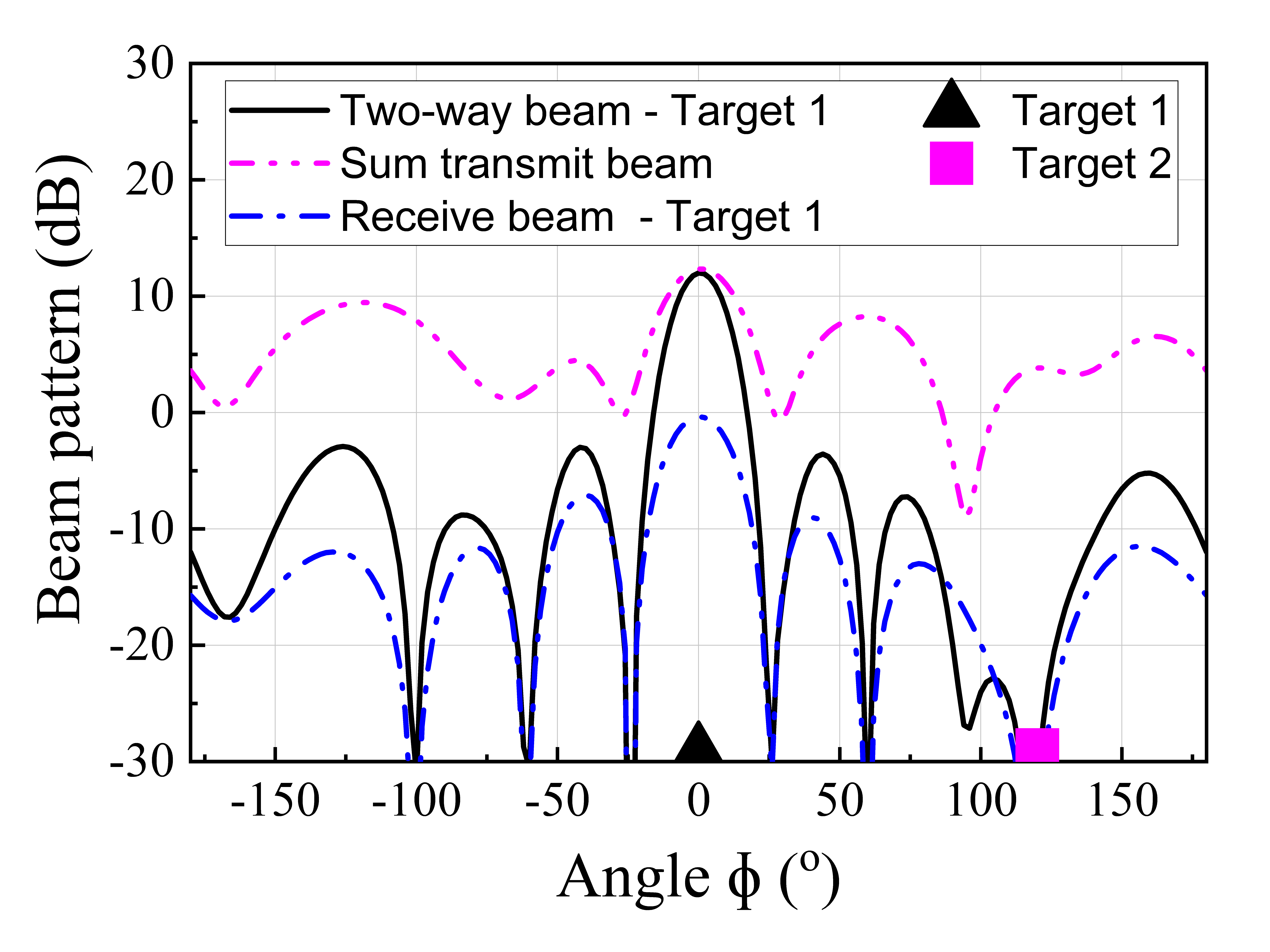}}
    \hfil
    \subfigure[Elevation plane $\phi = 0^\circ$]{\includegraphics[width=4.2 cm]{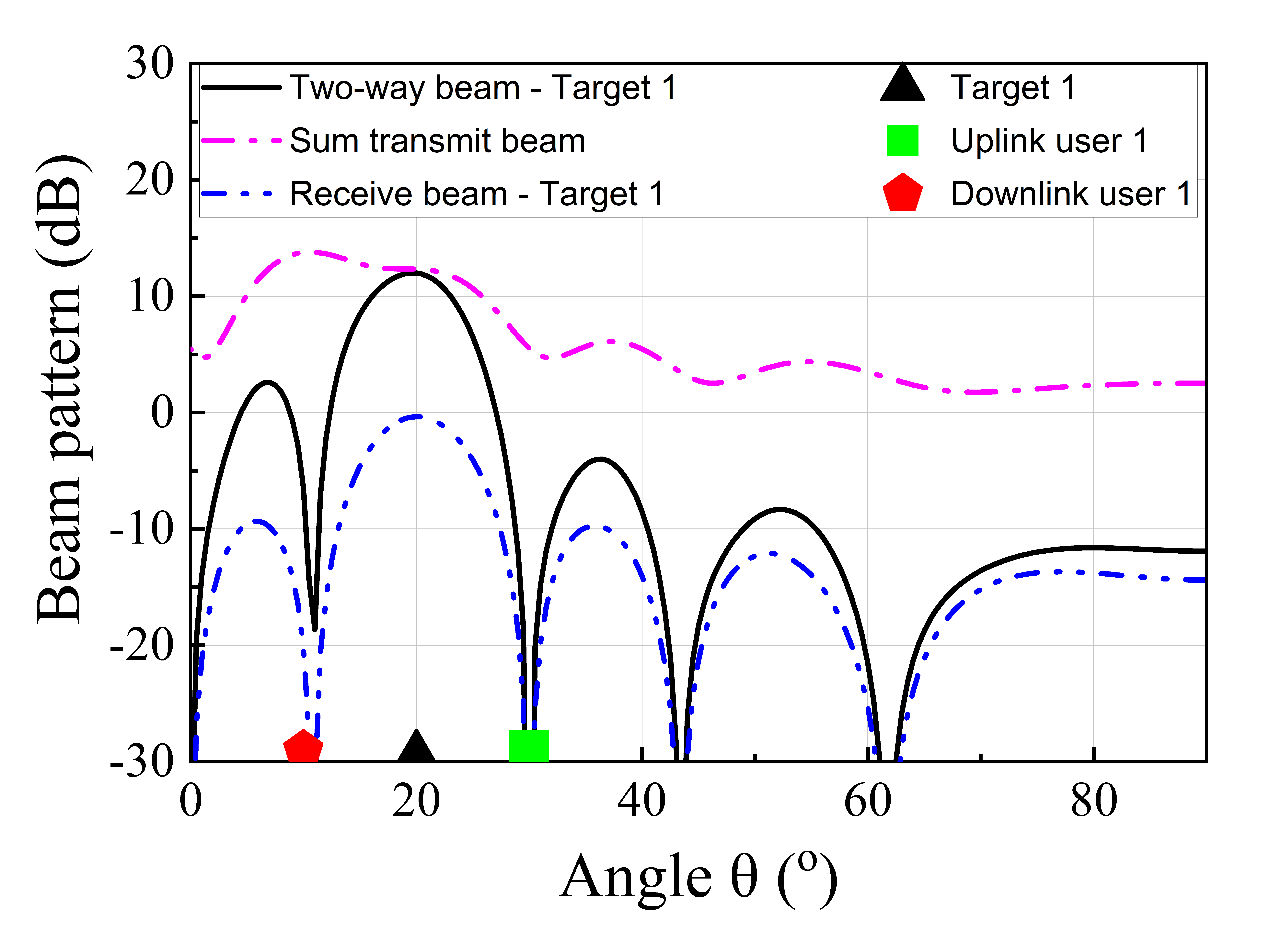}}
    \caption{Beam pattern of target 1 with required SINR $\mu_{\text{rad}}$ = 20 dB.}
    \vspace{-10pt}
    \label{beam_pattern2}
\end{figure}
As reported in \cite{b7}, a higher sensing SINR leads to an increased probability of detection, thereby improving the system's sensing performance. Fig. \ref{capacity} examines the sum rate of downlink and uplink communication versus the sensing SINR threshold $\mu_{\text{rad}}$, demonstrating the tradeoffs between communication performance and sensing performance. The sum rate of the Communication-Only baseline remains unchanged as it does not consider the sensing constraints, and thus serves as the upper bound for the ISAC communication performance. For all other algorithms, As $\mu_{\text{rad}}$ increases, the sum rates decrease, as higher sensing demands require more power to be allocated to sensing within the ISAC system, leading to reduced communication performance. In the scenario with 1 target, the proposed GRQ-GA algorithm nearly matches the same performance as the Communication-Only case and achieves a 98\% gain over the TDD HD scheme at SINR requirement of 14 dB, highlighting the advantages of the proposed FD ISAC system over its HD ISAC alternatives. Additionally, our proposed algorithm demonstrates a slightly higher performance compared to the SCA-SDR algorithm in the single-target scenario as the later suffers from approximation loss. Unlike SCA-SDR algorithm, which is limited to single-target applications, our algorithm is capable of handling multi-target scenarios, highlighting its broader applications. 

When investigating the effect of increasing number of targets while keeping the number of users fixed, it is observed that the sum rates for communication decrease, as expected. This decline is due to the additional power required for sensing to maintain the desired sensing performance. At a sensing requirement of 20 dB, corresponding to 99\% of detection rate \cite{b7}, increasing the number of targets from 1 to 2 and then to 3 results in a reduction in the sum rate by 3.3 bps/Hz and 9.1 bps/Hz, respectively. This provides valuable insights: as the number of targets increases, to maintain sensing SINR for all targets, the damage to the communication performance increases due to the inter-target interference. Preserving communication performance, on the other hand, requires adjustments to other parameters, such as increasing power or bandwidth.

Fig. \ref{beam_pattern1} and \ref{beam_pattern2} shows the transmit beam pattern, receive and two-way beam pattern gain of target 1's receiver  at different sensing SINR requirement of 14 dB and 20 dB in Azimuth plane and Elevation plane, which are defined by: 
\begin{equation}
\begin{aligned}
    &p_{tx}(\phi, \theta) = |\boldsymbol{a}^{(tx)}(\phi,\theta)\boldsymbol{x}|^2, \\
    &p_{1,rx}(\phi, \theta) = |\boldsymbol{w_{s,1}^*}^H\boldsymbol{a}^{(rx)}(\phi,\theta)|^2,\\
    &p_1(\phi, \theta) = |\boldsymbol{w_{s,1}^*}^H\boldsymbol{A}(\phi,\theta)\boldsymbol{x}|^2.
\end{aligned}
\end{equation}
Note that the transmit beam is the sum of individual transmit beams, and the receive beam for target 1 is extracted from the overall sum receive beam. First, we examine the transmit beam pattern. With a lower sensing SINR requirement of 14 dB, the beam prioritizes the downlink user to maximize the communication sum rate, as the beam peak aligning with the downlink user's position in the Elevation plane. Additionally, a slight notch is noticed in the transmit beam in the target's direction on the Elevation plane to mitigate some of the interference caused by the sensing target to the communication user. When the sensing SINR requirement is increased to 20 dB, we observe a stronger gain in the target's direction, as higher power is needed to meet the sensing's requirements.
Next, regarding the receive beam pattern, it can be seen that the peak is directed toward the desired target, while nulls are placed in the directions of other targets and users to suppress interference.
Finally, the two-way beam pattern, which combines the transmit and receive beams. It can be seen that the beamforming strategy was effectively put nulls in the users and other targets' direction to mitigate the interference. The peak gain toward the target depends on the sensing requirement. For the 14 dB requirement, the peak gain was recorded at 5.7 dB with a side-lobe level (SLL) of 6.8 dB, and for the 20 dB requirement, the peak gain and SLL increased to 12 dB and 15 dB in the Azimuth plane, respectively.

\section{Conclusion}\label{sec5}
In this paper, we examine joint beamforming optimization involving multiple users and multiple targets for an FD ISAC system. We formulate an optimization problem to maximize the total sum-rate of uplink and downlink communications while imposing SINR constraints for each target and user. First, we present a closed-form solution for obtaining the optimal receive beamformers using GRQ optimization, followed by solving the overall optimization using the floating-point GA approach. Numerical results show a 98\% improvement in sum-rate compared to the HD ISAC system, while achieving nearly the same performance as the communication-only system and having better performance compared to SDR-SCA algorithm in single target scenario. Additionally, when the number of targets increases, to maintain
sensing SINR for all targets, the damage to the communication performance increases due to the inter-target interference. Also, the impacts of sensing requirement to beam pattern are studied, showing that higher sensing requirement demands increased peak gain and SLL. Further work can be extended to cover frequency-selective channels in wideband and dynamic scenarios.
\section*{Acknowledgment}
The authors would like to acknowledge the financial support from the Dutch Research Council (NWO) under 3D-ComS project with file number 19751.  

\end{document}